\documentclass[acmsmall]{acmart}
\AtBeginDocument{%
  }

\setcopyright{acmlicensed}
\copyrightyear{2025}
\acmYear{2026}
\acmDOI{}
\acmISBN{978-1-4503-XXXX-X/2018/06}

\newcommand{\e}[1]{\mbox{\lstinline[basicstyle=\normalsize]|#1|}}
\usepackage{boogie}
\usepackage{eiffel}
\usepackage{listings}
\usepackage{graphicx}
\usepackage{url}
\usepackage{makecell}
\usepackage{paralist}
\usepackage{subfigure}
\usepackage{comment}
\usepackage{xcolor}

\usepackage{bbding}
\usepackage[]{graphicx}
\usepackage{scalerel}
\usepackage{algpseudocode}
\usepackage{algorithm}
\usepackage{amsmath}
\usepackage{colortbl}
\usepackage[]{hyperref}
\usepackage{soul}
\usepackage[export]{adjustbox}
\usepackage{wrapfig}

\makeatletter
\@addtoreset{chapter}{part}
\makeatother

\newcommand{\versiontwo}[1]{#1}
\newcommand{\versionthree}[1]{#1}

\begin{document}

\title{Combining Tests and Proofs \versionthree{} for Better Software Verification}


 \author{Li Huang}
 \affiliation{%
   \institution{Constructor University Bremen}
   \city{Bremen}
   \country{Germany}
 }
 \email{li.huang@constructor.org}

 \author{Bertrand Meyer}
 \affiliation{%
   \institution{Eiffel Software}
      \city{Santa Barbara}
   \country{USA}
 }
 \email{bm@construtor.org}

  \author{Manuel Oriol}
 \affiliation{%
   \institution{Constructor \versionthree{University}}
      \city{Bremen}
   \country{Germany}
 }
 \email{moriol@constructor.university}

\renewcommand{\shortauthors}{\versiontwo{Huang} et al.}

\begin{abstract}
Test or prove? These two approaches to software verification have long been presented as opposites. One is dynamic, the other static: a test executes the program, a proof only analyzes the program text. A different perspective is emerging, in which testing and proving are \versionthree{complementary rather than competing} techniques for producing software of verified quality.

Work performed over the past few years and reviewed here develops this complementarity by \versionthree{taking advantage of Design by Contract, as available in Eiffel}, and exploiting a feature of modern program-proving tools based on ``Satisfiability Modulo Theories'' (SMT): \textit{counterexample} generation. A counterexample is an input combination that makes the program fail. If we are trying to prove a program correct,  we hope not to find any. \versiontwo{One} can, however, apply counterexample generation to incorrect programs, as a tool for \textit{automatic test generation}. We can also introduce faults into a correct program and turn the counterexamples into an automatically generated \textit{regression test suite} with full coverage. Additionally, we can use these mechanisms to help produce \textit{program fixes} for incorrect programs, with a guarantee that the fixes are correct. All three applications\versionthree{, leveraging on the mechanisms of Eiffel and Design by Contract,} hold significant promise to address \versionthree{some of the} challenges of program testing, software maintenance and Automatic Program Repair.

\end{abstract}



\received{12 July 2025}

\maketitle
\section{From opposition to complementarity} \label{introduction}

Software development is not just an intellectual exercise but an engineering endeavor, \versiontwo{with  considerable benefits when it succeeds, but also the risk of considerable damage if the resulting software is incorrect. \textit{Software verification} seeks to avoid such adverse outcomes. It} has two variants, dynamic and static. Dynamic verification\versiontwo{, also known as testing,} consists of executing a system on sample input conditions and checking the results against expected properties (``oracles''). For many projects in industry, testing is the only form of verification. In contrast, ``verification'' among programming researchers usually refers to \textit{static} methods, which do not execute the program but analyze its text, usually in its source form; the analysis may be directed towards specific properties,
or it may perform a full proof of correctness, which assumes that developers have produced not only a program but also a formal specification of its intended behavior, and determines whether the program correctly implements that specification.

There is a long history of pitting tests against proofs. Practitioners often reject proofs as \versiontwo{too hard,} costly and impractical. \versiontwo{Critics of testing, for their part,} dismiss \versiontwo{them} as ineffective since they only exercise a minuscule subset of all possible cases, as famously stated by  Dijkstra: ``\textit{Program testing can be used to show the presence of bugs, but never to show their absence!}'' \cite{dijkstrastructured}. The controversy was at times heated, with  strong dismissals of proofs \cite{Goodenough}  (against Naur's proof attempts \cite{Naur}), \cite{DeMilloLipton}, and equally stark statements about the limitations of testing  by proponents of proofs.

In recent years, more conciliatory attitudes have developed, attested among others by the creation of the TAP conference, Tests And Proofs, running since 2007.
Verification is \versiontwo{difficult} in any \versiontwo{of its forms, and should not let any} dogma \versiontwo{exclude} any potentially useful technique. Even the staunchest proof advocate must  accept that it may not be possible to specify  all properties formally, particular ``non-functional'' attributes such as performance and fault tolerance, and that the value of any proof is conditional on assumption that the  compiler and hardware \versiontwo{both preserve program semantics,} leaving room for tests. One may also note the role of \textit{failures} on both sides:

\begin{itemize}
    \item A failed test is actually a proof: it demonstrates that the program is incorrect --- in the same way that in the Popperian view of science a negative experiment proves the theory wrong. (In this respect one may interpret Dijkstra's quip, usually heard as an indictment of tests, as \textit{advertisement }for them: it is one of the goals of software verification to help uncover faults.)

    \item A failed proof attempt often leaves us clueless at what is wrong. To understand, it is useful to see  a test triggering the underlying fault if any. Section \ref{proof2test} will develop this idea.
\end{itemize}

\noindent Test-and-proof complementarity goes beyond these observations. The catalyst for the work reported in this article is a feature of modern program proving tools, such as those based on SMT  (Satisfiability Modulo Theories)
\cite{barrett2010smt},
which try to generate \textit{counterexamples} violating the specification. A proof will succeed if it cannot produce any; but for an \textit{incorrect} program the counterexamples yield test\versiontwo{s}. \versiontwo{We} may also use the resulting information to propose corrections to the bugs.  For a correct program, we may obtain a regression test suite by injecting faults and producing counterexamples. 


We have applied these ideas to develop verification and program-repair techniques that \versionthree{exploit} the complementarity\versionthree{, while taking advantage of the Design by Contract approach to software construction \cite{meyer1997object}.} Described until now in separate articles \cite{huang2023failed,huang2022improving,huang2023seeding,huang2024execution,huang2024mcdc,huang2025loop}, \versiontwo{these steps} gain here a comprehensive and up-to-date presentation for a broader audience.
\versiontwo{Section \ref{core} explains the general idea: taking advantage of the information associated with proof failures. Section \ref{assertions} is a reminder on \versionthree{Design by Contract, }a key ingredient of the approach.} Subsequent sections introduce the successive refinements of the basic idea:
\begin{itemize}
    \item  Proof2Test (Section \ref{diagnosis} and \cite{huang2023failed}) turns failed proofs, often \versionthree{seemingly} unhelpful, into useful test cases, which offer concrete evidence of why the proofs fail, enabling programmers to understand and address the underlying issues. 
    \item
Proof2Fix (Section \ref{proof2fix} and \cite{huang2024execution}) extends the combination of tests and proofs to automatic program repair by providing a guaranteed validation of fixes through the prover.
\item 
`Seeding Contradiction'' (Section \ref{sc} and \cite{huang2023seeding}), 
extends the capability to systematic test generation. 
By deliberately creating failed proofs and running them through Proof2Test, it can produce a collection of tests that satisfy code coverage criteria. It also incorporates \textit{loop unrolling}
to create tests that explore \versiontwo{loop behavior} and helped provide an assessment of Modified Condition/Decision Coverage (MC/DC), a testing criterion sometimes viewed skeptically but widely used in industry.
\end{itemize}

\versiontwo{Section \ref{related} reviews earlier work on neighboring topics and Section \ref{conclusion} draws conclusions.}

\section {\versiontwo{The core idea: proofs and their failures}} \label{core}

\versiontwo{
The idea of proving programs correct (already prefigured in visionary late-1940s articles by Turing
and Von Neumann\versionthree{)}
goes back to the seminal work of Floyd, Hoare, Dijkstra, McCarthy, Naur and others in the 1970s and 1980s. What has changed in recent years is the appearance of practical \textit{tools}: ``provers'', which can prove the correctness of  programs, including large ones. Proving a significant program involves far too many steps to make human checking either practical or (if it somehow were possible) believable. The role of automation has consequently been essential. While still  used by only a minority  of  projects, provers have achieved some industry successes \cite{huang2023lessons}.

A prover works on the combination of a program and a specification (often expressed as ``contracts'' as explained in the next section).
The starting point for the present work is to take advantage of the technique used internally by many modern provers: they seek one or more ``counterexamples'' --- example inputs falsifying the expected properties. If this attempt \textit{fails}, the proof  \textit{succeeds}\footnote{The duality between failure and success is a general theme of the present work. It follows from the generalization of De Morgan's laws to predicate calculus: the negation of a universal is the existential of the negation, and conversely. In symbols, $\neg (\forall x \mid p) = \exists x \mid \neg p$. A proof seeks a universal ($\forall$) property; a failed test is an existential ($\exists$) property of the negation.}. The present work focuses on the intermediate steps, when the proof fails, meaning the search for a counterexample succeeds; and, more specifically on the counterexamples themselves.

Successive developments show that this idea has several fruitful applications. The first is to turn counterexamples into \textit{tests}. A proof failure is disconcerting: ``postcondition violated'' does not help the programmer understand what has been done wrong. Internally, though, the prover has a counterexample; ``for input 5, the output is -3 while required to be positive'' is considerably more informative for a programmer, even more so if accompanied by an automatically generated runnable test reproducing the failure. In other words, a tool can turn failed proofs into failed tests, with two major benefits: providing concrete, runnable evidence to the programmer; and \versionthree{yielding} a \textit{regression test} to be run again in the later history of the project once the bug has been fixed. Those are the functions of Proof2Test (Section \ref{proof2test}).

    The next application is to use the counterexamples to generate \textit{program fixes}. Automatic program repair (APR) needs ``invariants'' characterizing failed executions. Instead of getting them from manually generated tests, the present approach can generate guaranteed invariants; instead of validating fixes through tests, it can guarantee their correctness by running them through the prover. Those are the functions of Proof2Fix (Section \ref{proof2fix}).

For critical systems, the test suite \versionthree{must reach a high \textit{level of} \textit{coverage}, going beyond basic ``branch coverage'' to such industry-standard measures as} MC/DC\versionthree{, and \textit{unfolding} loops so that the tests can exercise several successive iterations rather than just zero or one (Section \ref{generation}).} 
}



\section{\versiontwo{The role of contracts}} \label{assertions}

While of general applicability, the results reported in this article rely \versiontwo{on a \textit{contract} mechanism \versionthree{as offered by various languages including the one used here, Eiffel.}
The need for contracts results from the observation that verifying the correctness of a program requires having a specification of its intended behavior. Contracts are elements of specification that are embedded in the code itself (rather than in separate documents) and, as a result, can be used for both testing (which will evaluate them in test runs) and proofs. They appear in  several modes:}
\begin{itemize}
    \item \versiontwo{Simple assertions (\e{check}, in Fig.~\ref{listing:simple code}): properties that must hold at specific points in the code.}
    \item Precondition (\e{require}, lines 3--4 in \versiontwo{Fig.~\ref{fig: proof-result-nocounterexample}}): properties that \emph{clients} must satisfy for any call.
    \item Postconditions (\e{ensure}, 18 -- 20 \versiontwo{in Fig.~\ref{fig: proof-result-nocounterexample}}): properties that the routine (\emph{supplier}) guarantees.
    \item Loop invariant (\e{invariant}, 9 -- 11 \versiontwo{in Fig.~\ref{fig: proof-result-nocounterexample}}): properties guaranteed after loop initialization and every iteration.
    \item Loop variant (\e{variant}, 16 \versiontwo{in Fig.~\ref{fig: proof-result-nocounterexample}}): integer measure that remains non-negative and decreases at each loop iteration, ensuring loop termination.
\end{itemize}

\noindent \versiontwo{As an example,} Fig. \ref{fig: proof-result-nocounterexample} shows an Eiffel class \e{MAX}.
The intent of the \e{max} function is to return into \e{Result} the maximum element of an integer array \e{a} of size \e{a.count}. The postcondition clauses in lines 19 and 20 (\emph{is\_max} and \emph{result\_in\_array}) specify the intent of the function: every element of the array should be less than or equal to \e{Result}; and at least one element should be equal to \e{Result}.

\begin{figure}[htbp]
\vspace{-0.3cm}
\centerline{{\includegraphics[width=11cm]{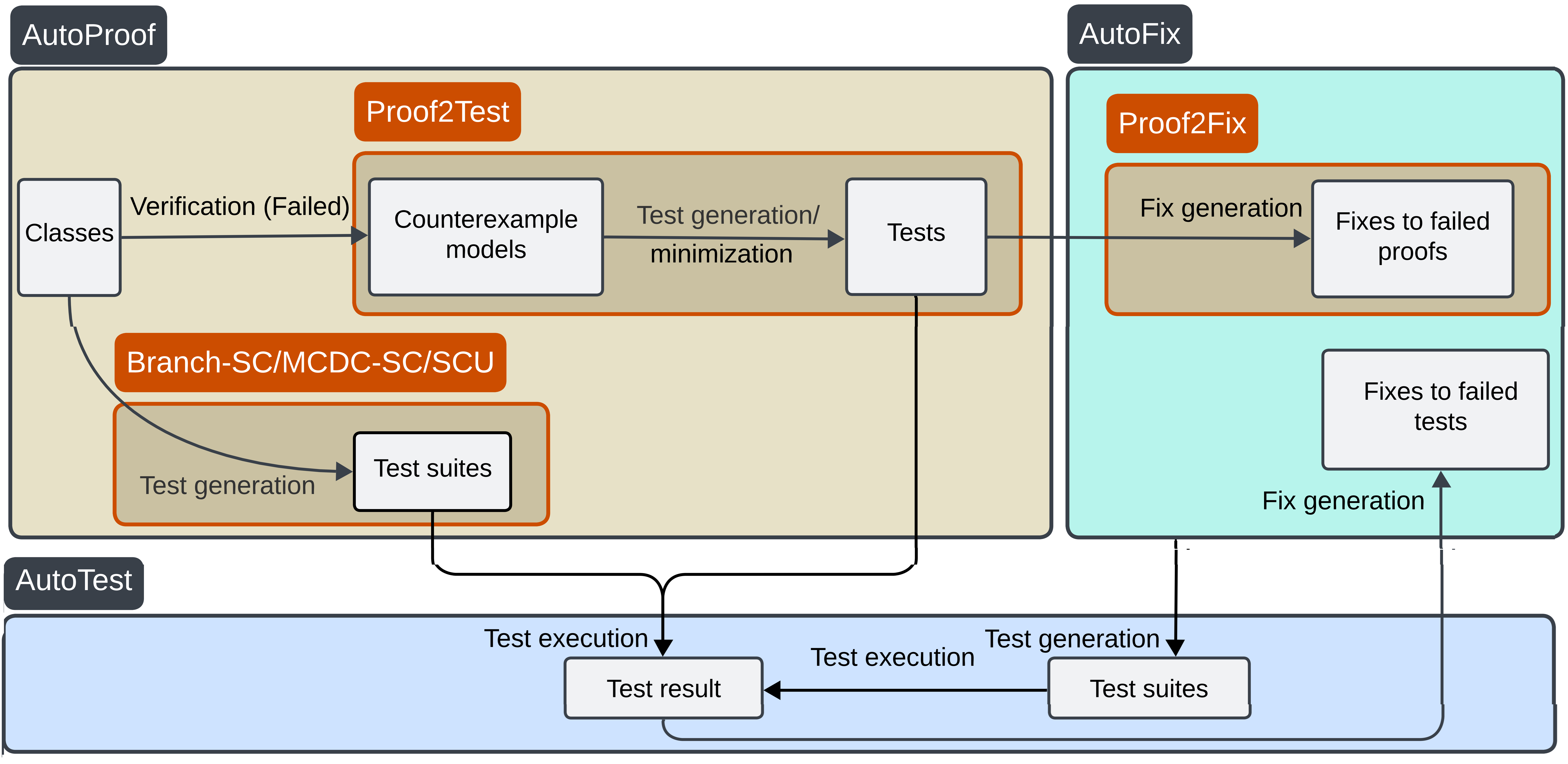}
}}
\caption{Combination of tools for program verification, testing and fixing}
\vspace{-0.3cm}
\label{fig:tool chain}
\end{figure}

\versiontwo{Contracts are a strong enabler for both testing and proving because they are specification elements embedded in the code itself. For testing, they make it possible to identify the source of a bug precisely: a precondition violation indicates a bug in the client; if the test passed the precondition, any subsequent failure is a bug in the code of the supplier method.
Tools \versionthree{such as AutoTest~\cite{meyer2007automatic}} have used these ideas to provide automated random testing. For proving, most tools rely on Dijkstra's weakest-precondition (wp) calculus~\cite{dijkstra1976discipline}, which \versionthree{deliver their full potential}  when \versionthree{the code already includes} postconditions and explicit loop variants and invariants.

Many programming languages (such as Java, C\#, C++, Python) support simple assertions but have no specific constructs for the other uses listed above. Extensions to some of them have introduced such constructs:   JML \cite{cok2021jml}
for Java and Spec\#
for C\#. A handful of languages, on the other hand, enjoy built-in contract mechanisms. They include  Eiffel,
SPARK,
Dafny
and the upcoming``C++26'' ISO/IEC C++ standard. The present work is based on Eiffel because of its extensive support for contracts as a core native mechanism rather than an add-on, its practical availability through tools such as the EiffelStudio compiler and IDE, and its long experience of research on tools using contracts for automatic test generation (AutoTest), automatic program repair (AutoFix) and proofs (AutoProof, used in the present work) \cite{meyer2007automatic, wei2010automated, tschannen2015autoproof}.}
All the tools and techniques described in this article have been integrated into the research version of EiffelStudio (Fig. \ref{fig:tool chain}).

\noindent \versiontwo{Contracts are also necessary to apply the theory used by many modern provers: ``Hoare logic'' \versionthree{of which} Dijkstra's weakest-precondition (wp) calculus \cite{dijkstra1976discipline} \versionthree{is a variant}. The basic condition to be established for a routine $r$ of body $b$, precondition $P$ and postcondition $Q$ is
\vspace{-0.2cm}
\[P \implies (b\ \textbf{wp}\ Q)\]
\noindent where $b$ \textbf{wp} $Q$ is the the weakest possible assertion such that $b$, started in a state satisfying $P$, will terminate in a state satisfying $Q$.
The prover will compute this condition for every $r$ and then prove that every call satisfies it. As discussed in the next section, provers based on SMT solving achieve this goal by trying to find an assignment of variables (a \textit{counterexample}) that contradicts it. The proof succeeds if they cannot. The present work exploits the converse case.

 }


\section{\versiontwo{Creating tests from failed proofs}} \label{diagnosis}
\label{proof2test}
\versiontwo{The observations of the previous section highlight} what a \textit{failure} means in the proving and testing processes, and how the two cases relate to each other. A failed proof leaves the programmer wondering what is wrong. A failed test provides concrete evidence.

\begin{figure}[htbp]
\vspace{-0.3cm}
\centerline{{\includegraphics[width=10cm]{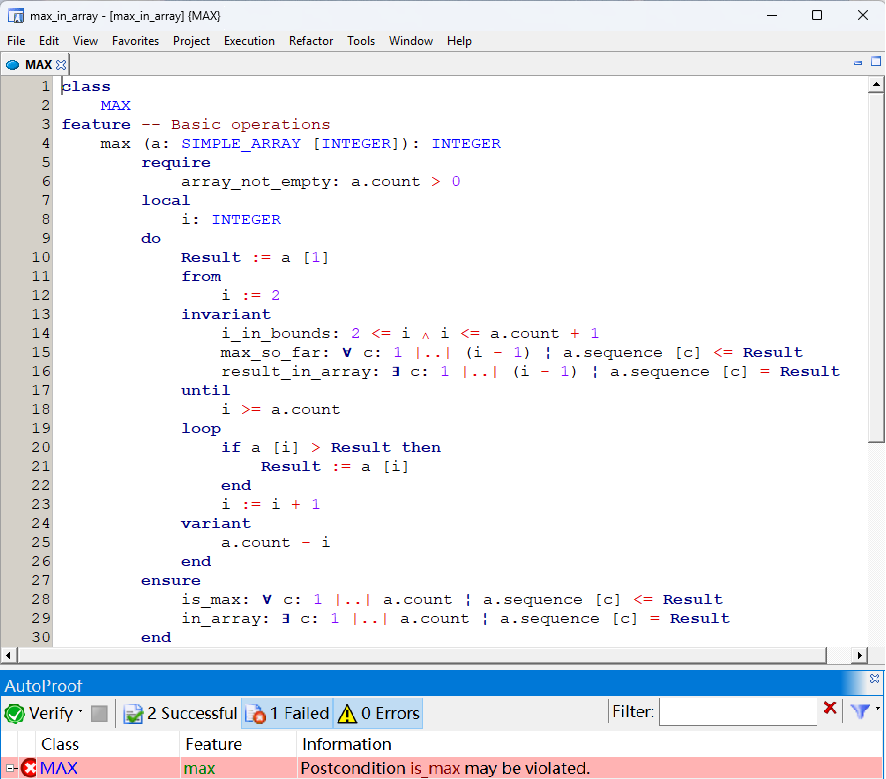}
}}
\vspace{-0.2cm}
\caption{A proof fails.}
\label{fig: proof-result-nocounterexample}
\vspace{-0.3cm}
\end{figure}

For anyone trying to prove a program correct, failure is a familiar companion. Textbooks usually show successful proofs, but in practice the path to success can be frustrating. You are using a prover and prepared the next proof step diligently: the program, its specification (contract\versionthree{s),} and extra annotations known as ``verification conditions''. You click "Verify", and the prover tells you it is not able to perform the proof successfully. Fig. \ref{fig: proof-result-nocounterexample} shows such a response, for the AutoProof prover,
before the inclusion of the tools discussed below.

\noindent Where \versiontwo{does one} go from here? The message states that a postcondition (line 28), describing the intended effect of the routine \emph{is\_max}, \textit{may be} violated. That does not even mean that the routine is incorrect; just that the prover cannot prove that it is correct. 
Is the implementation wrong (the most frequent case)? Is the \textit{specification} wrong? Did \versiontwo{the developer} forget to include some verification conditions? Or is the task exceeding the prover's abilities?

The process of getting to a successful proof has sometimes been called ``static debugging'', in reference to the classical form of debugging, which is ``dynamic'' as it involves running the program on tests. When they succeed, static techniques provide a guarantee of correctness that no  set of tests can \versiontwo{match}; but a failed test has the practical advantage of giving the programmer \versiontwo{ \textit{directly useful}} evidence of what is wrong, \versiontwo{helping to identify} the source of the failure --- \versiontwo{the bug (fault)}  --- and correcting it. \versiontwo{Such responses are} known in the psychology literature as constructive feedback~\cite {kluger1996effects}, although the basic idea is clear from daily-life experience (``\textit{your answer is wrong}'' is less effective than ``\textit{on page 3 you mistook pounds for kilograms}''). \versiontwo{In dealing with programming errors}, the programmer needs \versiontwo{mentally to reproduce}  the faulty program's execution. 
A failed test gives such a directly usable indication; the failed-proof message in Fig. \ref{fig: proof-result-nocounterexample} does not.

Modern proof tools have the potential to tell us more. \versiontwo{To establish Hoare-style properties as discussed in the previous section, AutoProof relies on} Boogie
\cite{le2011boogie}
which itself uses an ``SMT solver'' such as Z3 \cite{barrett2010smt}, whose method  to prove a property consists of looking for \textit{counterexamples} that defeat it. The proof succeeds when \versiontwo{it} cannot find any. \versiontwo{It fails if it finds one or more.}  \versiontwo{To gain more concrete information,} the programmer could then in principle \versiontwo{explore} the \versiontwo{solver's} internals \versiontwo{for} counterexamples. \versiontwo{Often, however, that} information is spread over hundreds of lines in internal files whose format, SMT-LIB \cite{barrett2010smt}, is intended for verification experts. Tools such as BVD (Boogie Verification Debugger)
and Boogaloo
can help but remain largely static.

\begin{figure}[htbp]
\vspace{-0.3cm}
\centerline{{\includegraphics[width=10cm]{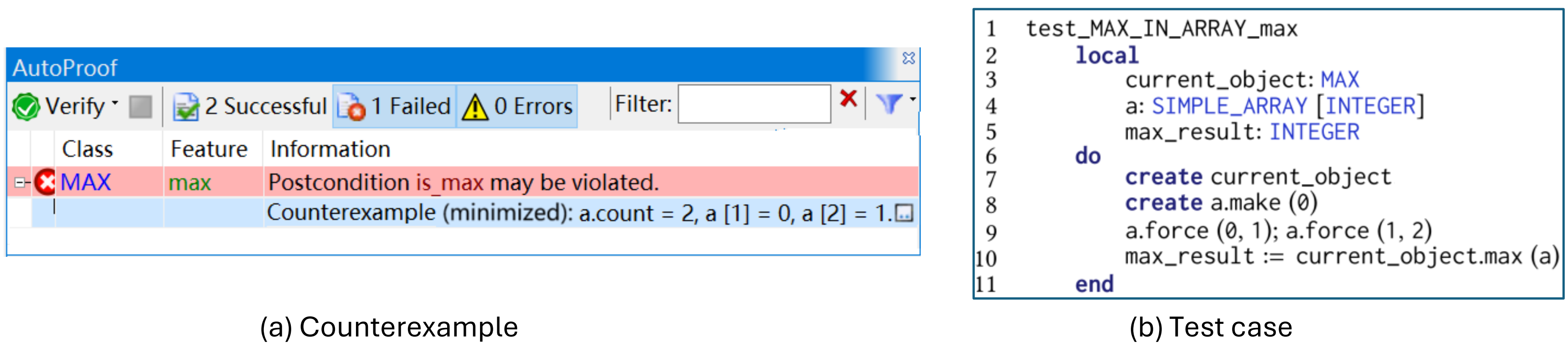}}}
\vspace{-0.2cm}
    \caption{\versiontwo{A failed proof providing a concrete (minimized) erroneous case}}
\vspace{-0.3cm}
\label{fig: proof2test-result}
\end{figure}

Proof2Test \cite{huang2023failed} \versiontwo{extends} AutoProof and provides the missing step. It analyzes internal information to display a directly understandable input example causing the failure, consisting here of the array's size and some of its elements, as shown in Fig. ~\ref{fig: proof2test-result} (a) \versiontwo{and then generates a test case as in Fig. ~\ref{fig: proof2test-result} (b)}.

\noindent In Fig. ~\ref{fig: proof2test-result} (a) the values are small: a two-element array with values 0 and 1. If asked to produce one counterexample, the SMT solver typically yields much larger values. \versiontwo{Here} it  \versionthree{initially} \versiontwo{produces} an array size \e{a.count} of 11,800 and values \e{a [11799] = 0}  and  \e{a [11800] = 5}. For disproof purposes (showing that the postcondition does not hold), \versiontwo{they} are as good as any others. Since the purpose of Proof2Test is to provide concrete evidence \versiontwo{helping} programmers figure out concretely what is going on, we prefer values that mean something at the human scale. Proof2Test takes advantage of the underlying prover's ability to generate not just one but several counterexamples (when provided with different \textit{seeds}). It applies a \textit{minimization} algorithm \cite{huang2022improving} to generate successive ones until it finds minimal values, such as those of Fig. ~\ref{fig: proof2test-result}. As with traditional programmer-devised test cases, they enable the programmer to relate to the failure by trying them out, as in Fig. \ref{fig: proof-result-nocounterexample}, and to see that on loop initialization at lines 10 -- 12 \e{Result} = 0 and \e{i} = 2 so that at line 18 the exit condition evaluates to \e{True}, terminating the loop and revealing the fault: early termination prevents the program from getting to the maximum value at position 2. To eliminate the error, it suffices to strengthen the exit condition to permit one more loop iteration: change \e{i} \e{$\geq$} \e{a.count} to \e{i} \e{$>$} \e{a.count}.


In addition to producing information that helps programmers understand and correct the fault behind a failure, Proof2Test can turn the counterexample into a \textit{test}. Fig. \ref{fig: proof2test-result} (b) is actually an executable test case produced by Proof2Test, \versionthree{expressed} in the \versionthree{programming} language. 
Programmers can run the test and observe, step by step, how it breaks the specification:
\begin{itemize}
\item Create an instance \e{current\_object} of class \e{MAX} (line 7).
\item Create an integer array \e{a} and fill it with values \e{0} at position \e{1} and \e{1} at index \e{2} (lines 8 -- 9).
\item Call the erroneous function \e{max} on \e{current\_object} with \e{a} as argument (line 10).
\end{itemize}



\noindent Running this code in AutoTest produces a run-time failure from the violation of the postcondition \emph{is\_max}, \versiontwo{providing} tangible evidence (not available from the failed proof attempt) of what is wrong with the version of Fig. \ref{fig: proof-result-nocounterexample} and making it possible to resume the development process.

A preliminary evaluation of Proof2Test on 20 Eiffel programs \cite{huang2023failed} demonstrated the approach's potential.  With an Intel 12-Core processor and 32 GB of RAM, the \versiontwo{above} process takes less than 0.5 seconds on average.
In most cases, the generated failing test is useful: executing it yields a specific trace illustrating how the program leads to the same contract violation that makes the proof fail; programmers can use the debugger's  step-by-step mode to understand the issue. 

In a small proportion of cases, the generated test does not lead to a failure. The reason is usually that the \textit{implementation} is  ``correct'' in some intuitive sense --- it does what the programmer somehow intended --- but the \textit{specification} is incomplete. These passing test runs are useful too, although in a different way from the failing ones: they alert the developers to a problem in the  specification. (We can hardly say that a program is correct if we are not able to state precisely \versionthree{what} it is supposed to do.) 
In either case, the generated tests are important as \textit{regression} tests: once the
 bugs have been corrected, every previously failing test should become part of the project’s
regression suite.

Experiment results \cite{huang2022improving} also show that minimization is cost-effective: in most cases, it reduces the values of integer
variables by over 80\% with an average cost of less than 4 extra reverification runs. Most minimized values are small and relatable: out of 125, 108 are in the range [-2, 2], out of which 58 are zero; others are usually close to some \versiontwo{constants} appearing in the program.


\section{\versiontwo{Combining tests and proofs for better program repair}} \label{repair}
\label{proof2fix}

Identifying  bugs is good; correcting them is better. Automatic program repair (APR) has demonstrated its potential in producing useful fixes. The Achilles' heel of most existing APR approaches~\cite{monperrus2018automatic}
is that they rely on test cases for both bug identification and fix validation. Writing test cases is tedious; running them can be time-consuming; and (the most significant limitation) validating them through tests does not guarantee their correctness (Dijkstra again). Proof2Fix \cite{huang2024execution}, based on Proof2Test, implements a static APR approach.

\begin{figure}[htbp]\vspace{-0.3cm}
\centerline{{\includegraphics[width=7.5cm]{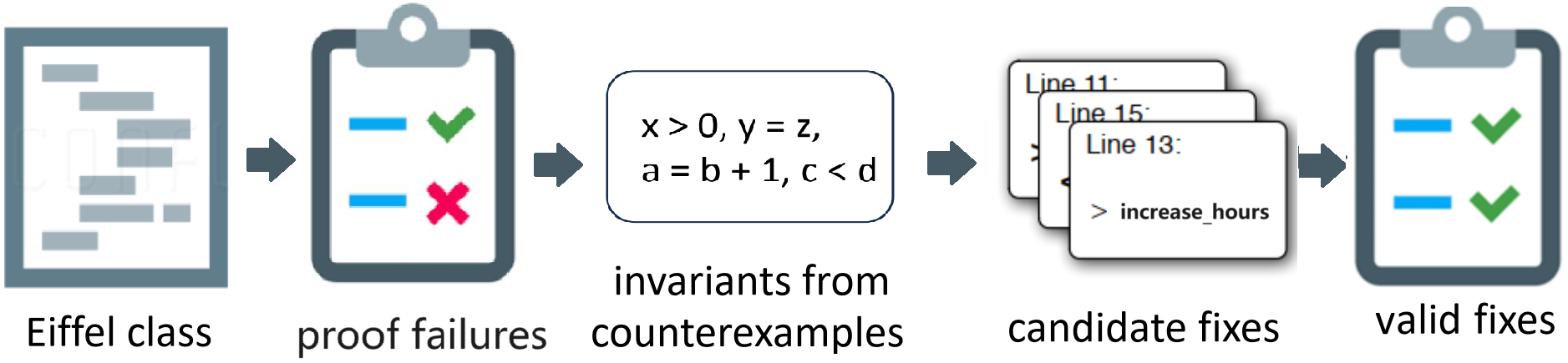}
}}
\vspace{-0.3cm}
\caption{Proof2Fix workflow}
\label{fig: workflow}
\end{figure}

The tool chain (Fig. \ref{fig: workflow}) is derived from the standard workflow in APR by replacing tests with proofs. The process starts with verifying a program using a prover (here AutoProof); if the prover detects a fault, Proof2Test generates counterexamples, each of which represents a failing execution trace, \versionthree{in} which the program goes to \versionthree{a} state that violates a desired property. Traditional APR techniques analyze failing \textit{tests} (usually generated manually) to identify any common elements which might pinpoint the cause of the failure;  Proof2Fix uses the same idea but with tests replaced by \textit{counterexamples} generated automatically and statically.



\vspace{-0.4cm}
\noindent The generation of abstract general properties from counterexamples uses Daikon \cite{ernst2007daikon}, an inference tool which produces a set of \emph{counterexample invariants}: predicates on the input variables which hold in the counterexamples, matched to a set of patterns (equality between variables or with constants, linear relations  \e{e1 = a $\cdot$ e2 + b} or \e{e1 =} \e{a $\cdot$ old e2 + b} etc.) 
The invariants make it possible to produce candidate fixes for either the contract or the implementation. (Empirical studies \cite{raluca} indicate that bugs arise from both kinds of mistake.) AutoProof then filters out invalid fixes, retaining only the fixes that pass the proof (\versionthree{they} remov\versionthree{e} the original proof failure and introduc\versionthree{e} no new one). Valid choices can then be presented to the programmer for decision.
Neither at the bug-finding stage nor at the bug-fixing stage is there any need to invent test data or to produce a test harness. 

Evaluation of fixing 80 proof failures shows \cite{huang2024execution} that the approach can produce meaningful fixes, formally validated.
It generates at least one valid fix for 82.5\% of all failures and at least one programmer-approved fix for 37.5\% of them.
A fixing session takes about 1 minute. 
Proof2Fix is most effective for failures caused by two types of faults: incorrect source expression in an assignment and incorrect condition in a conditional  instruction. In both cases, the inferred counterexample invariants correctly characterize the faulty cases that need to be ruled out. The approach is, on the other hand, not good at fixing failures caused by too-weak contracts, as the generated counterexamples exhibit too much diversity.


\section{\versiontwo{Proofs as a foundation for efficient test suite generation}} \label{generation}
\label{sc}
When the verification results in multiple failures, Proof2Test produces a number of different tests.
\versionthree{The} Seeding Contradiction \versionthree{strategy} (SC) \cite{huang2023seeding}, takes advantage of this possibility.

Every serious software project needs a regression test suite, essential to \versiontwo{managing} the project's evolution by ensuring that new developments do not invalidate previously working functionality. (Software engineering history includes numerous examples of regression bugs, such as the 2012 Knight Trading Group bug,
which lost the firm \$440 million in 45 minutes and almost bankrupted it, and the July 2024 CrowdStrike Falcon Sensor update.)
Aside from the practical advice of including a test for every case that failed at some point in the project's history, it is very difficult to produce a regression test suite with a high coverage of all the possible execution paths.

At first sight, the Proof2Test techniques described above seem inapplicable, since they deal with faulty programs, whereas we need a regression test suite for a working program. The idea behind Seeding Contradiction is, as the name suggests, that we will make a correct program incorrect --- in many different ways --- by inserting (``seeding'') \versiontwo{a bug} in every \versiontwo{branch}. While counterintuitive at first (the usual goal in software engineering is to go from incorrect to correct!), this idea enables us to benefit from the preceding techniques, which generate counterexamples for faulty programs. Under suitable conditions, a \textit{counterexample} for an \textit{incorrect} program can be an \textit{example} (a case of correct behavior) for the corresponding \textit{correct} program.  

\begin{wrapfigure}{r}{5cm}
\vspace{-0.3cm}
\centerline{{\includegraphics[width=5cm]{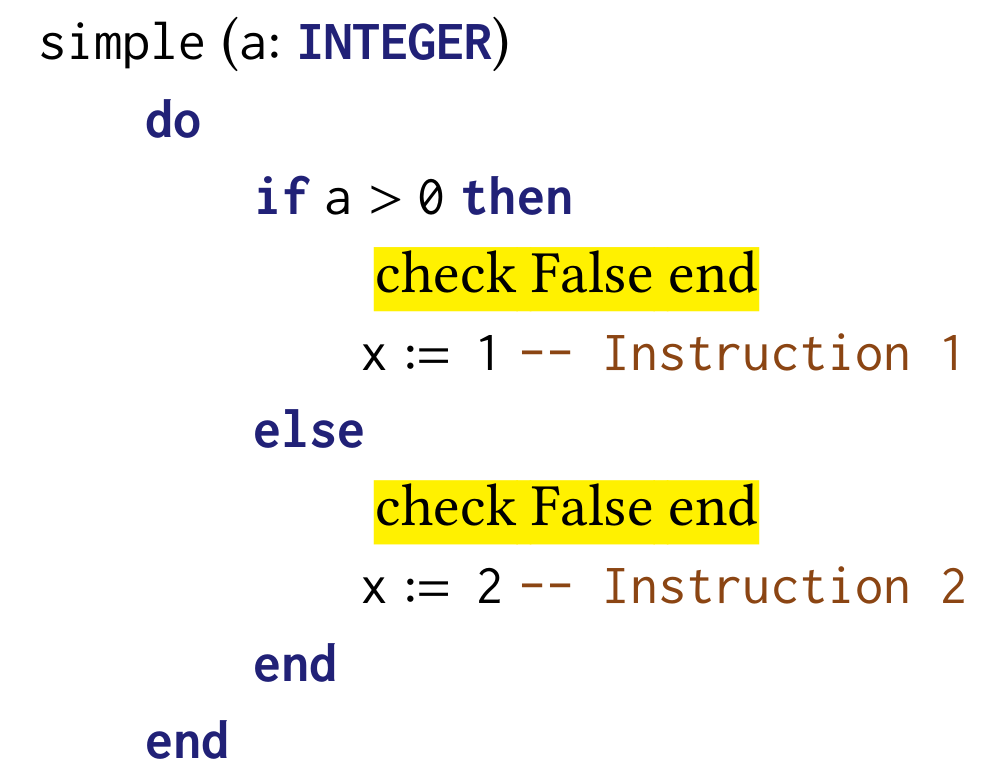}}
}
    \caption{A two-branch program with seeded contradictions}
     \label{listing:simple code}
\vspace{-0.3cm}
\end{wrapfigure}

Proof2Test, with Seeding Contradiction, inserts a faulty instruction (``trap property''
\cite{gargantini1999using}),
into every control branch, as shown in Fig. \ref{listing:simple code}. The instruction  is a
special case of the instruction
\e{check p end}
which states that the boolean property \e{p} must hold at the corresponding program point, causing the verification to fail if it cannot guarantee it. (Execution, if attempted, would \versionthree{cause an exception} if \e{p} does not hold at run time.) With \e{False} for \e{p}, the branch will fail to verify and the Proof2Test mechanisms of the previous section will generate the corresponding counterexample. The technique actually works for an  incorrect program as well as for a correct one. Even though it uses a prover normally intended for programs equipped with contract \versionthree{elements} (preconditions, postconditions, class \versionthree{and loop} invariants, \versionthree{loop variants}), it is applicable \versionthree{to} uncontracted programs as in this example.


By construction, the generated test suite achieves 100\% \textit{branch coverage} for \versiontwo{feasible} branches (excluding unreachable \versiontwo{ones}, which no test can possibly exercise). Proof2Test goes beyond branch coverage in two complementary ways:
\begin{itemize}
    \item MC/DC. On option, Proof2Test enforces the ``Modified Condition/Decision Coverage'' criterion,  recommended in several important industrial software safety standards 
    particularly observed in the aerospace industry.\footnote{\versiontwo{Examples are DO-178B (Software Considerations in Airborne Systems and Equipment Certification) from the Radio Technical Commission for Aeronautics and  ISO 26262-3:2011 (Road Vehicles - Functional Safety).}} 
    
    \item Loop unrolling. The body of a standard ``while'' or ``until'' loop, by definition, can be executed a variable number  of times; but branch coverage only needs a path with zero execution and another with at least one execution. In other words, it treats a loop like a conditional, missing bugs that \versionthree{arise} only with other numbers of execution. Proof2Test includes a loop unrolling mechanism, which considers loop bodies executed (``unrolled'') any number of times, up to a user-settable maximum determined in consideration of performance constraints. 
\end{itemize}

\noindent Our measurements  of the results of enforcing MC/DC  and loop unrolling \cite{huang2025loop} indicate that both enhancements uncover bugs that plain branch coverage misses. The example programs, while significant and in some cases sophisticated (coming for example from verification competitions) are still small, so we refrain from sweeping conclusions, but the results are encouraging:

\begin{itemize}
    \item MC/DC increases the number of found bugs by 12.2\% over branch coverage, growing to 14.6\% for programs with complex multiple-condition decisions \cite{huang2024mcdc}. This \versionthree{result} is significant since there is a certain skepticism towards MC/DC in the testing community (due in part to the original papers' lack of empirical results); some authors even dismiss the approach \cite{mcdcparadox}.
    
    \item Loop unrolling uncovers more faults as the depth of unrolling increases. Measurements show an improvement of 14.5\% over plain branch coverage with 5-level unrolling, uncovering 86.1\% of all faults~\cite{huang2025loop}. The rise of detected faults is steep for small unrolling levels ($\leq$ 5); going from 1 to 2 brings a 9\% improvement. Beyond 5, the effect  decreases (by 3\%  from 5 to 8). 
\end{itemize}

\noindent Many advanced test generation techniques, such as directed automated random testing
and concolic testing
\cite{sen2005cute},
still require some code execution. The techniques described here are entirely static: they perform verification, using a prover, but do not execute the program and hence do not require test data (although they \textit{generate} such data) or test harnesses. As noted, they work on correct as well as incorrect programs. 
If a program calls a routine with a precondition, the generated tests will satisfy the preconditions thanks to SC's constraint-solving techniques.

The results summarized above, from empirical studies \cite{huang2023seeding, huang2024mcdc, huang2025loop}, are achieved with reasonable performance. The SC strategy produces a test suite much faster than some of the dynamic techniques, such as IntelliTest \cite{tillmann2008pex} (previously known as Pex, a symbolic execution test-generation tool for .NET)  and AutoTest,
a test generation tool for Eiffel using Adaptive Random Testing, specifically ARTOO \cite{ciupa2008artoo}. For a typical small program, SC requires approximately 0.5 seconds to generate a 100\%-branch-coverage test suite; that  is about 50 times faster than IntelliTest and 500 times faster than AutoTest. Adding MC/DC multiplies the generation time by about 2 and the test suite's size (an important concern in practice since a good software process runs test suites frequently) by a factor of less than 10. Adding unrolling for test generation, depending on the complexity of the loop's control-flow structures, would lead to different increment on generation time. 
For plain loops (with neither embedded conditionals nor nested loops), it will lead to a roughly linear growth of generation time as the unrolling depth increases.
While for nested loops or loops with more complex conditionals embedded inside, as more failing instructions are introduced during test generation, the increment of generation time becomes more substantial. The evaluation in \cite{huang2025loop} shows that for the programs with nested loops, the time increases exponentially and becomes unacceptably high (over an hour) when the unrolling depths reaches 8.

\vspace {-0.2cm}
\section{Related work} \label{related}

A number of previous or parallel efforts turn counterexamples, generated from failed verification, into diagnosis messages. The Boogie Verification Debugger (BVD) \cite{le2011boogie} allows programmers, statically, to  ``debug'' Boogie proof-failure reports statically \cite{le2011boogie}. Other tools \cite{chakarov2022better, muller2011using} pursue similar goals. These approaches provide static traces for analysis; the approach reported here (Section \ref{diagnosis}) goes further by producing a dynamic trace and actual tests that programmers can run to see concretely what is going on. 

Many testing projects have used SMT solving techniques to generate tests. Some, such as Klee \cite{cadar2021klee}, PathCrawler \cite{ williams2005pathcrawler}, Pex/IntelliTest \cite{tillmann2008pex} and CUTE \cite{sen2005cute} are based on symbolic execution.
They explore paths in a program and use a constraint solver to reason about their feasibility. 
These strategies include a dynamic component and as a result cannot provide guarantees of exhaustive branch coverage. The testing strategies reported in the present work (Section \ref{generation}) are static. 


\versiontwo{The earliest work we know to have applied this idea \cite{angeletti2009improving}
generates tests for low-level C programs using Bounded Model Checking (BMC) \cite{kroening2014cbmc}, producing test suites with exhaustive branch coverage. A more recent variant, for Java bytecode, is JBMC \cite{brenguier2023jbmc}, where each verification run  only activates one assertion at a time, producing one counterexample (the
 C version \cite{angeletti2009improving} uses compile-time macros, one for each block to avoid actually generating multiple programs); in contrast, the SC framework produces only one program, using
a single \textit{run-time} variable representing the block number. \versionthree{Similarly, \cite{beyer2004generating} extends the software model checker Blast to generate test suites of full coverage with respect to given predicates.} BMC-based approaches rely on the correctness of the \textit{bound} of the execution trace: if the bound is not set correctly, some branches might not be covered, requiring more verification runs to obtain a better bound. The SC approach does not need such mechanisms.
An approach~\cite{nilizadeh2022generating} that applies ideas of fault injection for generating tests (building on work using counterexamples for program \versionthree{repair~\cite{nilizadeh2021more})} in  Hoare-style verification exploits counterexamples produced by the OpenJML \cite{cok2021jml}  verification tool to generate unit tests in JUnit format. These approaches all bear similarities to the present work, which seems to provide the most far-ranging and unifying collection of proof-test combination techniques, from exploiting proof failures to counterexample minimization, test generation covering several code coverage measures and automatic program repair. 
}

\versiontwo{Automated test generation is an active research area (e.g. \cite{10988978}) which in recent years has increasingly used LLM, generally leading to average coverage \versionthree{levels of} between 75\% to 90\%.
Proof2Test is much more efficient. LLMs can also serve to generate specifications automatically \cite{11029962}, although questions remain: while such specifications might make the code provable, they might also be the wrong specifications, preventing the generation of some relevant tests.} 

\versiontwo{The present work distinguishes \versionthree{itself} by the use of Eiffel. While \versionthree{it can be} viewed as a limitation, \versionthree{this choice} also provides a considerable advantage since it does not require an addition to another language (as JML for Java) but uses an existing (ISO-standardized) language, taking advantage of the built-in contract mechanism and hence directly using the original contract-equipped source code for both tests (with EiffelStudio's mechanism for evaluating assertions during testing) and proofs. Users of other languages may view the present results as those of ``laboratory work'', using ideal conditions \versionthree{so as }to open the way for others to extend the results to other environments. }

\vspace{-0.2cm}
\section{Conclusion\versiontwo{s}} \label{conclusion}
\label{conclusions}

The work presented here suffers from a number of limitations. Examples so far are still small, \versionthree{although some} involve sophisticated algorithms. Some of the applications (but not all) assume that the programs have been equipped with contracts, a step that not all programmers are prepared to take. 
The program-repair applications so far have mostly covered variables of basic types and need to be extended to complex data structures.

These results, however, are promising, in particular on the performance side: verification is faster than running many tests, \versionthree{an} advantage \versionthree{which} grows rapidly with the program's size and complexity. More generally, any arguments along the lines that programmers allegedly do not want to write contracts should be weighed against the tasks that \versionthree{programmers} have to carry out in today's dominant test-based approach. The commonly used phrase ``automatic testing'', in its current meaning, is a misnomer, obscuring the unpleasant reality:
the enormous burden that test case generation, still largely a manual task, imposes on programmers. (AI tools will help, but they will also help writing contracts.) With a static approach, that task can be automated.

More generally, program proofs and dynamic tests pursue the same ultimate goal of software correctness. This article has shown that one can treat them as complementary rather than exclusive. We have seen three major applications of the idea, exploiting the features of SMT solvers underlying modern proof tools: turning failed proofs, often elusive to the programmer, into directly usable tests  (especially after \textit{minimization}) evidencing the failure, which brings examples to human scale; generating 100\%-coverage test suites entirely automatically; and providing guaranteed bug fixes. Today's tools are still imperfect but they point the way to a major leap in the effectiveness of software verification and repair, combining the best of what the software engineering community has learned both on the side of tests and on the side of proofs.



\bibliographystyle{splncs04}
\bibliography{reference}

@INPROCEEDINGS {10988978,
author = { Gorla, Daniele and Kumar, Shivam and Roselli Lorenzini, Pietro Nicolaus and Alipourfaz, Alireza },
booktitle = { 2025 IEEE Conference on Software Testing, Verification and Validation (ICST) },
title = {{ CubeTesterAI: Automated JUnit Test Generation Using the LLaMA Model}},
year = {2025},
volume = {},
ISSN = {2159-4848},
pages = {565-576},
doi = {10.1109/ICST62969.2025.10988978},
url = {https://doi.ieeecomputersociety.org/10.1109/ICST62969.2025.10988978},
publisher = {IEEE Computer Society},
address = {Los Alamitos, CA, USA},
month =apr}

@INPROCEEDINGS {11029962,
author = { Ma, Lezhi and Liu, Shangqing and Li, Yi and Xie, Xiaofei and Bu, Lei },
booktitle = { 2025 IEEE/ACM 47th International Conference on Software Engineering (ICSE) },
title = {{ SpecGen: Automated Generation of Formal Program Specifications via Large Language Models }},
year = {2025},
volume = {},
ISSN = {},
pages = {16-28},
doi = {10.1109/ICSE55347.2025.00129},
url = {https://doi.ieeecomputersociety.org/10.1109/ICSE55347.2025.00129},
publisher = {IEEE Computer Society},
address = {Los Alamitos, CA, USA},
month =May}

@inproceedings{wei2010automated,
  title={Automated fixing of programs with contracts},
  author={Wei, Yi and Pei, Yu and Furia, Carlo A and Silva, Lucas S and Buchholz, Stefan and Meyer, Bertrand and Zeller, Andreas},
  booktitle={Proceedings of the 19th international symposium on Software testing and analysis},
  pages={61--72},
  year={2010}
}

@inproceedings{huang2024execution,
  title={Execution-Free Program Repair},
  author={Huang, Li and Meyer, Bertrand and Mustafin, Ilgiz and Oriol, Manuel},
  booktitle={Companion Proceedings of the 32nd ACM International Conference on the Foundations of Software Engineering},
  pages={517--521},
  year={2024}
}

@inproceedings{huang2024mcdc,
  title={Is MCDC Really Better? Lessons from Combining Tests and Proofs},
  author={Huang, Li and Meyer, Bertrand and Oriol, Manuel},
  booktitle={International Conference on Tests and Proofs},
  pages={25--44},
  year={2024},
  organization={Springer}
}

@inproceedings{huang2025loop,
  author       = {Li Huang and Bertrand Meyer and Reto Weber},
  title        = {Loop Unrolling: Formal Definition and Application to Testing},
  booktitle    = {Proceedings of the International Conference on Testing Software and Systems (ICTSS 2025)},
  year         = {2025},
  publisher    = {Springer Lecture Notes in Computer Science},
  note        = {Preprint available as arXiv:2502.15535}
}

@article{huang2023failed,
  title={A failed proof can yield a useful test},
  author={Huang, Li and Meyer, Bertrand},
  journal={Software Testing, Verification and Reliability},
  volume={33},
  number={7},
  pages={e1859},
  year={2023},
  publisher={Wiley Online Library}
}

@inproceedings{raluca,
  title={Systematic evaluation of test failure results},
  author={Meyer, Bertrand and Ciupa, Ilinca and Liu, Lisa Ling and Oriol, Manuel and Leitner, Andreas and Borca-Muresan, Raluca},
  booktitle={Workshop on Reliability Analysis of System Failure Data (RAF)},
  year={2007}
}

@inproceedings{tillmann2008pex,
  title={{Pex--White Box Test Generation for  .Net}},
  author={Tillmann, Nikolai and De Halleux, Jonathan},
  booktitle={International Conference on Tests and Proofs (TAP)},
  pages={134--153},
  year={2008},
  organization={Springer}
}

@inproceedings{huang2023seeding,
  title={Seeding Contradiction: a fast method for generating full-coverage test suites},
  author={Huang, Li and Meyer, Bertrand and Oriol, Manuel},
  booktitle={IFIP International Conference on Testing Software and Systems},
  pages={52--70},
  year={2023},
  organization={Springer}
}

@article{gargantini1999using,
  title={Using model checking to generate tests from requirements specifications},
  author={Gargantini, Angelo and Heitmeyer, Constance},
  journal={ACM SIGSOFT Software Engineering Notes},
  volume={24},
  number={6},
  pages={146--162},
  year={1999},
  publisher={ACM New York, NY, USA}
}

@inproceedings{huang2022improving,
  title={{Improving Counterexample Quality from Failed Program Verification}},
  author={Huang, Li and Meyer, Bertrand and Oriol, Manuel},
  booktitle={International Symposium on Software Reliability Engineering Workshops (ISSREW)},
  pages={268--273},
  year={2022},
  organization={IEEE}
}

@article{huang2023lessons,
  title={Lessons from Formally Verified Deployed Software Systems},
  author={Huang, Li and Ebersold, Sophie and Kogtenkov, Alexander and Meyer, Bertrand and Liu, Yinling},
  journal={ACM Computing Surveys, to appear},
  year={2025},
  note={Preprint (of extended version) available as arXiv:2301.02206}
}

@article{monperrus2018automatic,
  title={Automatic software repair: A bibliography},
  author={Monperrus, Martin},
  journal={ACM Computing Surveys (CSUR)},
  volume={51},
  number={1},
  pages={1--24},
  year={2018},
  publisher={ACM New York, NY, USA}
}

@incollection{dijkstrastructured,
  author    = {Edsger W. Dijkstra},
  title     = {Notes on Structured Programming},
  booktitle = {Structured Programming},
  editor    = {O.-J. Dahl and E. W. Dijkstra and C. A. R. Hoare},
  pages     = {1--82},
  publisher = {Academic Press},
  year      = {1972},
  note      = {Originally circulated in 1970},
}

@inproceedings{meyer2007automatic,
  title={Automatic testing of object-oriented software},
  author={Meyer, Bertrand and Ciupa, Ilinca and Leitner, Andreas and Liu, Lisa Ling},
  booktitle={SOFSEM 2007: Theory and Practice of Computer Science: 33rd Conference on Current Trends in Theory and Practice of Computer Science, Harrachov, Czech Republic, January 20-26, 2007. Proceedings 33},
  pages={114--129},
  year={2007},
  organization={Springer}
}

@article{brenguier2023jbmc,
  title={Jbmc: A bounded model checking tool for java bytecode},
  author={Brenguier, Romain and Cordeiro, Lucas and Kroening, Daniel and Schrammel, Peter},
  journal={arXiv preprint arXiv:2302.02381},
  year={2023}
}

@inproceedings{ciupa2008artoo,
  title={ARTOO: adaptive random testing for object-oriented software},
  author={Ciupa, Ilinca and Leitner, Andreas and Oriol, Manuel and Meyer, Bertrand},
  booktitle={Proceedings of the 30th international conference on Software engineering},
  pages={71--80},
  year={2008}
}

@inproceedings{kroening2014cbmc,
  title={CBMC--C Bounded Model Checker (Competition Contribution)},
  author={Kroening, Daniel and Tautschnig, Michael},
  booktitle={Tools and Algorithms for the Construction and Analysis of Systems: 20th International Conference, TACAS 2014, Held as Part of the European Joint Conferences on Theory and Practice of Software, ETAPS 2014, Grenoble, France, April 5-13, 2014. Proceedings 20},
  pages={389--391},
  year={2014},
  organization={Springer}
}

@article{cadar2021klee,
  title={KLEE symbolic execution engine in 2019},
  author={Cadar, Cristian and Nowack, Martin},
  journal={International Journal on Software Tools for Technology Transfer},
  volume={23},
  pages={867--870},
  year={2021},
  publisher={Springer}
}

@article{sen2005cute,
  title={CUTE: A concolic unit testing engine for C},
  author={Sen, Koushik and Marinov, Darko and Agha, Gul},
  journal={ACM SIGSOFT Software Engineering Notes},
  volume={30},
  number={5},
  pages={263--272},
  year={2005},
  publisher={ACM New York, NY, USA}
}

@inproceedings{angeletti2009improving,
  title={Improving the Automatic Test Generation process for Coverage Analysis using CBMC.},
  author={Angeletti, Damiano and Giunchiglia, Enrico and Narizzano, Massimo and Palma, Gabriele and Puddu, Alessandra and Sabina, Salvatore},
  booktitle={RCRA@ AI* IA},
  year={2009}
}

@inproceedings{williams2005pathcrawler,
  title={Pathcrawler: Automatic generation of path tests by combining static and dynamic analysis},
  author={Williams, Nicky and Marre, Bruno and Mouy, Patricia and Roger, Muriel},
  booktitle={European Dependable Computing Conference},
  pages={281--292},
  year={2005},
  organization={Springer}
}

@inproceedings{nilizadeh2021more,
  title={{More Reliable Test Suites for Dynamic APR by Using Counterexamples}},
  author={Nilizadeh, Amirfarhad and Calvo, Marlon and Leavens, Gary T and Le, Xuan-Bach D},
  booktitle={International Symposium on Software Reliability Engineering (ISSRE)},
  pages={208 -- 219},
  year={2021},
  organization={IEEE}
}

@inproceedings{nilizadeh2022generating,
  title={{Generating Counterexamples in the Form of Unit Tests from Hoare-style Verification Attempts}},
  author={Nilizadeh, Amirfarhad and Calvo, Marlon and Leavens, Gary T and Cok, David R},
  booktitle={International Conference on Formal Methods in Software Engineering (FormaliSE)},
  pages={124--128},
  year={2022},
  organization={IEEE}
}

@inproceedings{chakarov2022better,
  title={{Better Counterexamples for Dafny}},
  author={Chakarov, Aleksandar and Fedchin, Aleksandr and Rakamari{\'c}, Zvonimir and Rungta, Neha},
  booktitle={International Conference on Tools and Algorithms for the Construction and Analysis of Systems (TACAS)},
  pages={404--411},
  year={2022},
  organization={Springer}
}

@inproceedings{muller2011using,
  title={{Using Debuggers to Understand Failed Verification Attempts}},
  author={M{\"u}ller, Peter and Ruskiewicz, Joseph N},
  booktitle={International Symposium on Formal Methods (FM)},
  pages={73--87},
  year={2011},
  organization={Springer}
}

@inproceedings{barrett2010smt,
  title={{The SMT-LIB Standard: Version 2.0}},
  author={Barrett, Clark and Stump, Aaron and Tinelli, Cesare and others},
  booktitle={{International Workshop on Satisfiability Modulo Theories}},
  volume={13},
  pages={14},
  year={2010}
}

@book{dijkstra1976discipline,
  title={{A Discipline of Programming}},
  author={Dijkstra, Edsger Wybe},
  year={1976},
  publisher={Prentice Hall}
}

@inproceedings{le2011boogie,
  title={{The Boogie Verification Debugger}},
  author={Le Goues, Claire and Leino, K Rustan M and Moskal, Micha{\l}},
  booktitle={International Conference on Software Engineering and Formal Methods (SEFM)},
  pages={407--414},
  year={2011},
  organization={Springer}
}

@book{meyer1997object,
  title={{Object-Oriented Software Construction, second edition}},
  author={Meyer, Bertrand},
  year={1997},
  publisher={Prentice Hall}
}

@inproceedings{tschannen2015autoproof,
  title={{Autoproof: Auto-active Functional Verification of Object-Oriented Programs}},
  author={Tschannen, Julian and Furia, Carlo A and Nordio, Martin and Polikarpova, Nadia},
  booktitle={International Conference on Tools and Algorithms for the Construction and Analysis of Systems (TACAS)},
  pages={566--580},
  year={2015},
  organization={Springer}
}

@inproceedings{beyer2004generating,
  title={{Generating Tests from Counterexamples}},
  author={Beyer, Dirk and Chlipala, Adam J and Henzinger, Thomas A and Jhala, Ranjit and Majumdar, Rupak},
  booktitle={International Conference on Software Engineering (ICSE)},
  pages={326--335},
  year={2004},
  organization={IEEE}
}

@inproceedings{cok2021jml,
  title={{JML and OpenJML for Java 16}},
  author={Cok, David R},
  booktitle={International Workshop on Formal Techniques for Java-like Programs (FTfJP)},
  pages={65--67},
  year={2021},
  publisher={ACM}
}

@article{ernst2007daikon,
  title={The Daikon system for dynamic detection of likely invariants},
  author={Ernst, Michael D and Perkins, Jeff H and Guo, Philip J and McCamant, Stephen and Pacheco, Carlos and Tschantz, Matthew S and Xiao, Chen},
  journal={Science of computer programming},
  volume={69},
  number={1-3},
  pages={35--45},
  year={2007},
  publisher={Elsevier}
}

@inproceedings{Goodenough,
  author    = {John B. Goodenough and Susan L. Gerhart},
  title     = {Towards a Theory of Test Data Selection},
  booktitle = {Proceedings of the International Conference on Reliable Software},
  pages     = {493--510},
  year      = {1975},
  publisher = {ACM},
  address   = {New York, NY, USA},
  doi       = {10.1145/800027.808444}
}

@article{Naur,
  author  = {Peter Naur},
  title   = {Programming by Action Clusters},
  journal = {BIT Numerical Mathematics},
  volume  = {9},
  number  = {3},
  pages   = {250--261},
  year    = {1969},
  publisher = {Springer},
  doi     = {10.1007/BF01955237}
}

@article{DeMilloLipton,
  author    = {Richard A. DeMillo and Richard J. Lipton and Alan J. Perlis},
  title     = {Social Processes and Proofs of Theorems and Programs},
  journal   = {Communications of the ACM},
  volume    = {22},
  number    = {5},
  pages     = {271--280},
  year      = {1979},
  publisher = {ACM},
  doi       = {10.1145/359104.359106}
}

@article{kluger1996effects,
  author       = {Avraham N. Kluger and Angelo DeNisi},
  title        = {The effects of feedback interventions on performance: A historical review, a meta-analysis, and a preliminary feedback intervention theory},
  journal      = {Psychological Bulletin},
  volume       = {119},
  number       = {2},
  pages        = {254--284},
  year         = {1996},
  publisher    = {American Psychological Association},
  doi          = {10.1037/0033-2909.119.2.254}
}

@article{mcdcparadox,
  title={The MCDC paradox},
  author={Bhansali, Praful V},
  journal={ACM SIGSOFT Software Engineering Notes},
  volume={32},
  number={3},
  pages={1--4},
  year={2007},
  publisher={ACM New York, NY, USA}
}

@String{BIT = "{BIT}" }

@String{Computing = "Computing" }

@String{Computer = "{IEEE} Computer" }

@String{Academic = "Academic Press" }

@String{Springer = "Springer-Verlag" }

@BOOK{test,
   author = "Donald E. Knuth",
   title = "Seminumerical Algorithms",
   volume = 2,
   series = "The Art of Computer Programming",
   publisher = "Addison-Wesley",
   address = "Reading, MA",
   edition = "2nd",
   month = "10~" # jan,
   year = "1981",
}

@ArtifactSoftware{R,
    title = {R: A Language and Environment for Statistical Computing},
    author = {{R Core Team}},
    organization = {R Foundation for Statistical Computing},
    address = {Vienna, Austria},
    year = {2019},
    url = {https://www.R-project.org/},
}

\end{document}